\documentclass[aps,prb,twocolumn,preprintnumbers,amsmath,amssymb,nofootinbib,floatfix]{revtex4}
\pdfoutput=1
\usepackage{graphicx}		
\usepackage{dcolumn}		
\usepackage{bm}			

\begin{document}

\title{Effect of hybridization on structural and magnetic properties of iron-based superconductors}
\author{R. A. Jishi}
\author{H. M. Alyahyaei}
\affiliation{ Department of Physics, California State University, Los Angeles, California 90032 }

\date{\today}

\begin{abstract}
 We show that the strong hybridization between the iron 3d
and the arsenic 4p orbitals, in the newly discovered iron-based high-T$_{c}$ superconductors, leads to an explanation
of certain experimental observations that are presently not well understood. The existence of a lattice distortion, the smallness
of the Fe magnetic moment in the undoped systems, and the suppression of both the lattice distortion and the magnetic order
upon doping with fluorine, are all shown to result from this hybridization.

\end{abstract}
\maketitle

	An interetsing development occurred recently with the discovery of a new class of layered, iron-based,
high temperature superconductors. Kamihara et al.~\cite{Kamihara:2008}  reported a superconducting transition temperature T$_{c}$=26 K in
fluorine doped LaOFeAs, and T$_{c}$ increased under pressure to 43 K.~\cite{Takahashi:2008} 
Replacement of lanthanum with other rare earth elements gave a series of superconducting compounds ReO$_{1-x}$F$_{x}$FeAs where
Re = Ce, Pr, Nd, or Sm, with transition temperatures close to or exceeding 50 K.~\cite{Zocco:2008, Ren:2008, Chen:2008, x_Chen:2008}. 
Oxygen-deficient samples were also fabricated and found to superconduct at 55 K.~\cite{Ren_2:2008} 
The parent compound, ReOFeAs, has a tetragonal unit cell with two ReOFeAs molecules, and it
consists of a stack of alternating ReO and FeAs layers. 
 The FeAs layer, in reality, consists of a square planar sheet of Fe atoms sandwiched between two sheets
of As atoms, such that each Fe atom is tetrahedrally coordinated with four As atoms.

	Certain facts that emerged from neutron scattering measurements are not well understood:

1) A tetragonal-to-orthorhobic structural phase transition takes place in the undoped system at T$_{S}$$\sim$150 K, below which an
 antiferromagnetic (AFM) order is established.~\cite{Cruz:2008,Zhao:2008,Qiu:2008,Zhao_j:2008}
It was suggested~\cite{Yildirim:2008} that the tetragonal-to-orthorhombic distortion results from magnetic frustration. However, the magnetic
order occurs more than 20 K below T$_{S}$ in LaOFeAs and PrOFeAS,~\cite{Cruz:2008,Dong:2008}
casting doubt on such an interpretation.

2) The measured magnetic moment of Fe is very small, $\sim$0.4-0.8 $\mu_{B}$,~\cite{Cruz:2008,Zhao:2008,Qiu:2008,Zhao_j:2008} whereas
 calculations based on density functional
theory (DFT) give $\sim$2 $\mu_{B}$.~\cite{Ishibashi:2008,Dong:2008,Ma:2008,Cao:2008}
   
3) It is observed that upon replacing a small fraction of the oxygen atoms in the ReO layer by fluorine atoms, the
tetragonal-to-orthorhombic distortion is gradually suppressed, along with the AFM order.~\cite{Cruz:2008,Zhao:2008,Qiu:2008,Zhao_j:2008}

	In this letter we present a scheme within which the above observations may be explained. It is based on
the consequences of the strong hybridization between the Fe 3d and the As 4p orbitals. On the one hand, this hybridization gives
rise to superexchange between different Fe atoms, leading to the establishment of the AFM order. On the other
hand, the hybridization will have a strong effect on the states of any given Fe atom; it is this effect that we explore here, and
show that it offers an explanation of the above observations.
\begin{figure}
   \includegraphics[width=0.45\textwidth]{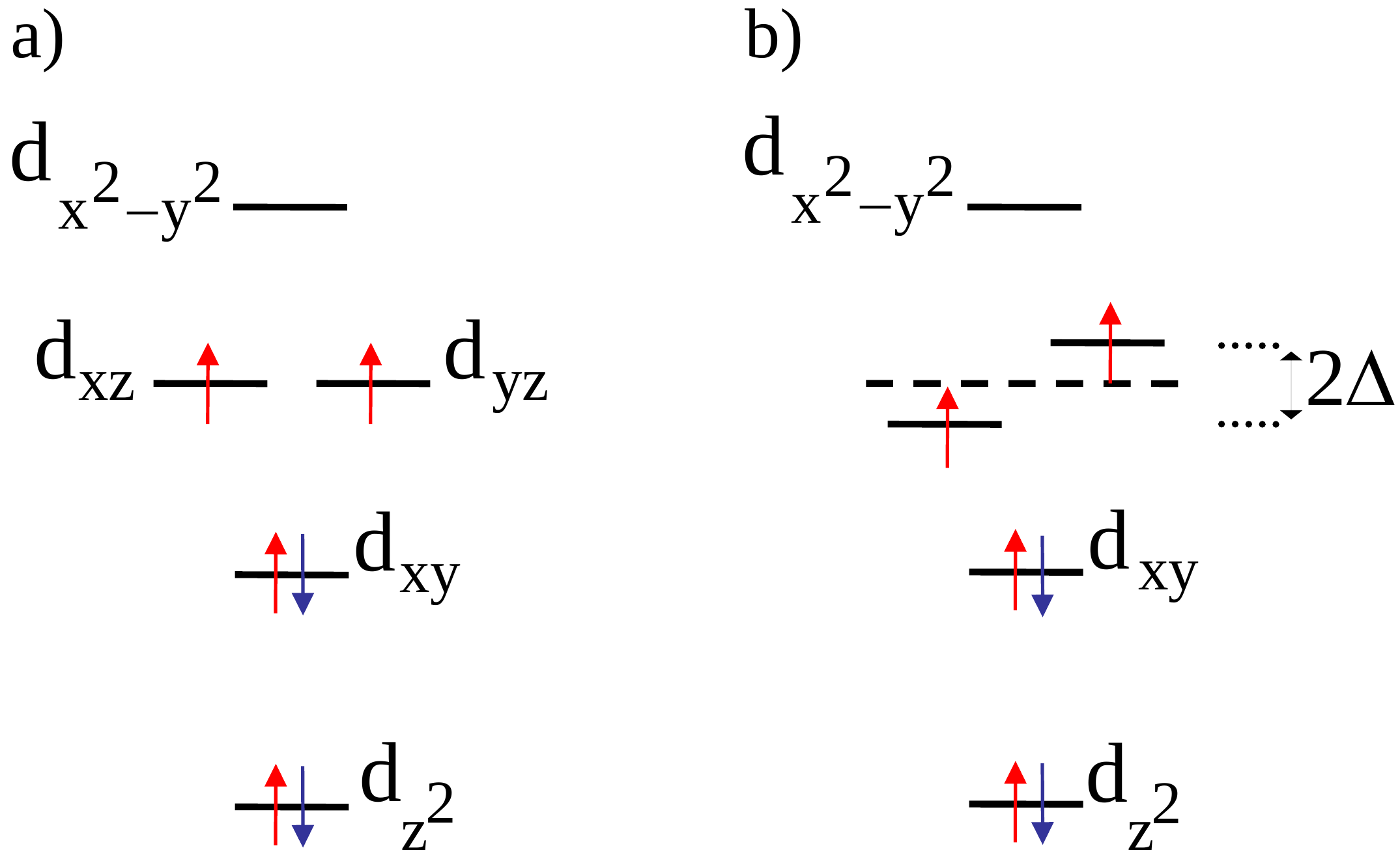}
   \caption{\label{fig:figure1} The occupation scheme of the Fe 3d levels in the a) tetragonal and b) orthorhombic structures of ReOFeAs, where 
Re is a lanthanide.}
\end{figure}

In the tetragonal phase, and in the absence of hybridization, the crystal field at a given Fe site, produced by the four surrounding
As ions, has $S_{4}$ symmetry, and it splits the otherwise degenerate Fe 3d levels into three 
nondegenerate levels and one doubly degenerate level~\cite{Cao:2008} as shown in Fig.~\ref{fig:figure1}a. Since DFT calculations indicate that an Fe site
has a total spin 1, the six 3d electrons are taken to occupy orbitals as shown in Fig.~\ref{fig:figure1}; this occupation scheme
 was also suggested by Li~\cite{Tao:2008} and Baskaran~\cite{Baskaran:2008}.
Following Li,~\cite{Tao:2008} we assume that the energy splitting produced by the crystal field is sufficiently large to neglect the filled
3d$_{z^{2}}$ and 3d$_{xy}$ orbitals and the empty 3d$_{x^{2}-y^{2}}$ orbital in discussing the low energy behavior. 
 A tetragonal-to-orthorhombic lattice distortion will lift 
the remaining degeneracy, and assuming that the lattice constant $a$ is larger than $b$, the energy level scheme of the d orbitals will be as shown in
Fig.~\ref{fig:figure1}b. If in Fig.~\ref{fig:figure1}a we set $\epsilon_{d_{xz}} = \epsilon_{d_{yz}} = \epsilon$,
then in Fig.~\ref{fig:figure1}b we have 
$\epsilon_{d_{xz}} = \epsilon - \Delta \equiv \epsilon_{1}$ and $ \epsilon_{d_{yz}} = \epsilon + \Delta \equiv \epsilon_{2}$. 
We consider a system comprising an Fe atom and an adjacent As atom, for which the Hamiltonian may be expressed as 
\[
\mathcal{H} = \mathcal{H}_{0} + \mathcal{H}^{'},   ~ \mathcal{H}_{0} = \mathcal{H}_{d} + \mathcal{H}_{p},
\]
\[
\mathcal{H}^{'} =  \sum_{\sigma} \alpha c_{p\sigma}^{\dag} c_{d_{xz,\sigma}} + \sum_{\sigma} \beta c_{p\sigma}^{\dag} c_{d_{yz,\sigma}} + H.C.
\]
Here $\mathcal{H}_{d}$ and $\mathcal{H}_{p}$ are the Hamiltonians for the Fe and As atoms, respectively, $c_{p\sigma}^{\dag}$ $(c_{p\sigma})$ is the 
creation (annihilation) operator for an electron in the As 4p orbital with spin $\sigma$, \(c_{{d_{xz}},\sigma}^{\dag}\)\((c_{{d_{xz}},\sigma})\) creates 
(annihilates) an electron with spin $\sigma$ in the Fe 3d$_{xz}$ orbital, with a simillar definition applying to  \(c_{{d_{yz}},\sigma}^{\dag}\)\((c_{{d_{yz}},\sigma})\), 
$\alpha$ and $\beta$ are hopping matrix elements, taken for simplicity to be constant and real, and H.C. stands for hermitian conjugates
 of the last two terms in 
$\mathcal{H}^{'}$. The terms in $\mathcal{H}^{'}$ express the hybridization between the Fe 3d orbitals and the As 4p orbitals, 
in the form of electron hopping between these orbitals. 
\begin{figure}
   \includegraphics[width=0.5\textwidth]{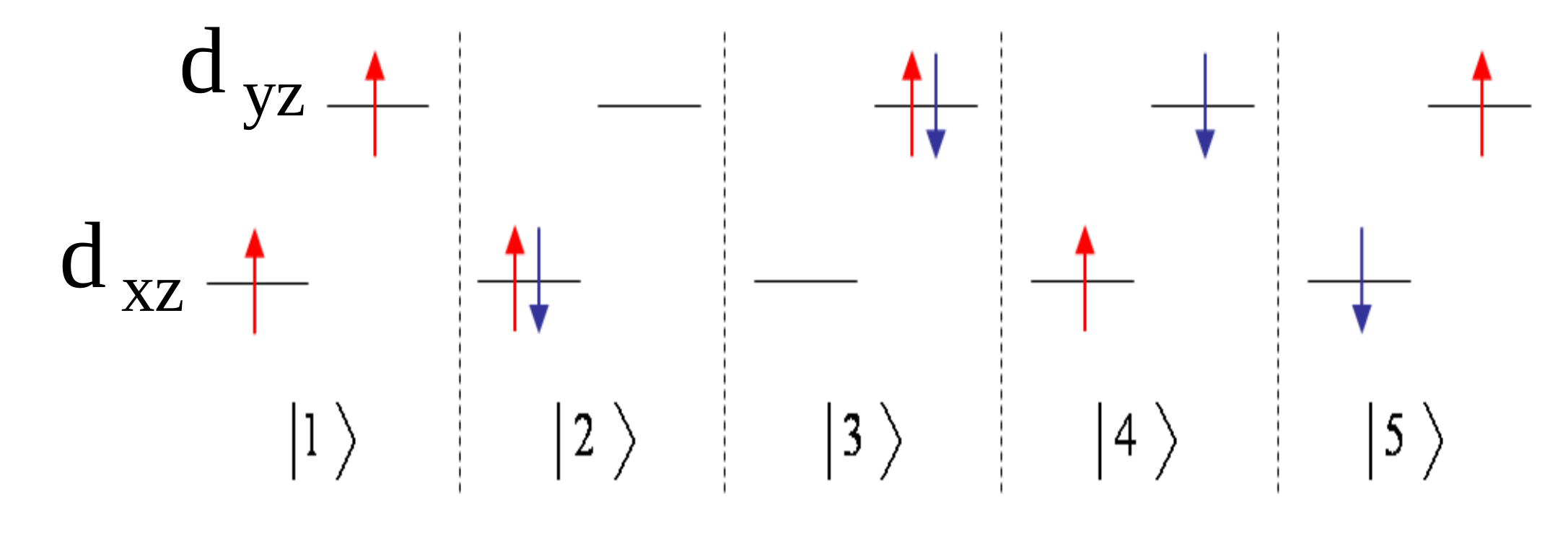}
   \caption{\label{fig:figure2} The state $|1\rangle$ along with the other four states that can be reached from it through hybridization of
the Fe 3d and the As 4p orbitals.}
\end{figure}

The states shown in Fig.~\ref{fig:figure1} are eigenstates of $\mathcal{H}_{0}$, but not $\mathcal{H}$.  
 Nonvanishing matrix elements connect the state $|1\rangle$ to other states shown in Fig.~\ref{fig:figure2}.  
State $|2\rangle$ can be reached from $|1\rangle$ by the spin-up electron in Fe 3d$_{yz}$ hopping to the As ion, followed by a spin-down electron 
hopping from the As ion to the Fe 3d$_{xz}$ orbital. Alternatively, a spin-down electron may hop first from
 the As ion to the Fe 3d$_{xz}$ 
orbital, followed by the electron in the Fe 3d$_{yz}$ orbital hopping to the As ion. The ground state of the Fe atom is now obtaind by diagnolizing 
the Hamiltonian matrix in the subspace spanned by these five states. 

 The Hamiltonian is projected onto this 5-dimensional subspace by a canonical transformation 
\begin{eqnarray*}
H_{T} &=& e^{-S} H e^{S}\\ 
      &=& H + [H,S] + \frac{1}{2} [[H,S],S] + ...\\
      &=& H_{0} + H^{'} + [H_{0},S] + [H^{'},S] + \frac{1}{2} [[H_{0},S],S].     
\end{eqnarray*}
Choosing the operator S such that $[H_{0},S] = -H^{'}$, and keeping terms to 2$^{nd}$ order in H$^{'}$, we obtain 
\[
H_{T} = H_{0} + \mathcal{H}_{eff}
\]
where $\mathcal{H}_{eff} = \frac{1}{2} [H^{'},S]$. The matrix elements of H$_{eff}^{'}$, connecting eigenstates of H$_{0}$, are given by 
\begin{eqnarray*}
\langle n| H_{eff}^{'} |m\rangle &=& \frac{1}{2} \langle n| H^{'}S - SH^{'}|m\rangle\\ 
                                 &=& \frac{1}{2} \sum_{k} [\langle n|H^{'}|k\rangle \langle k|S|m\rangle - \langle n| S|k\rangle \langle k|H^{'}|m\rangle]
\end{eqnarray*}
Using $H^{'} = - [H_{0} , S]$. we obtain 
\[
\langle n|H_{eff}^{'}|m\rangle = \frac{1}{2} \sum_{k} (\frac{1}{E_{m}-E_{k}} + \frac{1}{E_{n}-E_{k}}) H_{nk}^{'} H_{km}^{'} 
\]
The eigenstates $|m\rangle$ of H$_{0}$ are in fact composite states comprising the 3d$_{xz}$ and 3d$_{yz}$ orbitals of the Fe atom, along with 
the 4p
orbitals of As. We carried out DFT calculations on undoped LaOFeAs, along with Bader's atoms in molecules (AIM) theory,~\cite{Bader:1990} and found that  
As has a charge of $-0.92e$ and Fe has a charge of $0.30e$, where $e$ is the charge of the proton. The charge transferred from LaO to 
the FeAs layer resides 
mainly on the As 4p$_{z}$ orbital, which has an energy much lower than that of the As 4p$_{x}$ or 4p$_{y}$ orbital. Furthermore, by examining
 the projected density of states (PDOS) reported 
by Cao et al.~\cite{Cao:2008} we see that the DOS at the Fermi level is dominated by the Fe 3d$_{xz}$ and 3d$_{yz}$ states, while the DOS of the p
 states has 3 peaks: one peak, which results from filled 4p$_{z}$ bands, is 3-4 eV below the Fermi energy E$_{f}$,
 and 2 other peaks, one being about 0.5 eV below and the other 1 eV 
above E$_{F}$; these two peaks result from the 4p$_{x}$ and 4p$_{y}$ orbitals. Since there are two As atoms per unit cell, we can
assume that there are two occupied 4p$_{z}$ bands, 3-4 eV below E$_{F}$, two occupied bands with mixed 4p$_{x}$ and 4p$_{y}$ character, $\sim$0.5 eV
below E$_{F}$, and another two empty bands with mixed 4p$_{x}$ and 4p$_{y}$ character, $\sim$1 eV above E$_{F}$. Considering any
single As ion, we may assume that it has four electrons in the 4p orbitals, two of which occupy the 4p$_{z}$ orbital, and the other two
electrons occupy an orbital with mixed 4p$_{x}$ and 4p$_{y}$ character which, for convenience, we simply call p, and there is an empty
orbital, also with mixed 4p$_{x}$ and 4p$_{y}$ character which we call p'.
We ignore the filled, low energy 4p$_{z}$ orbital
, and we describe the state of the As atom by the occupation of the p and p$^{'}$ orbitals.
 The eigenstate $|1\rangle$ of H$_{0}$ is
 written explicitly as 
\[
|1\rangle = |Fe; d_{xz,\uparrow} , d_{yz, \uparrow}\rangle |As; p_{\uparrow \downarrow} \rangle .  
\]  
 The eigenstate $|2\rangle$ of H$_{0}$ is 
\[
|2\rangle = |Fe;  d_{xz, \uparrow \downarrow}\rangle |As; p_{\uparrow},p_{\uparrow}^{'} \rangle 
\]
Note that $|2\rangle$  differs from $|1\rangle$ by having both d electrons in the d$_{xz}$ orbital and by a spin flip in the As p orbitals. 
In order to construct the matrix element $\langle 2|H_{eff}^{'}|1\rangle$ we need the intermediate  states $|k\rangle $. As discussed earliear there are two 
such states, 
\[
|k_{1}\rangle = |Fe; d_{xz,\uparrow} \rangle |As; p\uparrow \downarrow, p^{'}\uparrow \rangle,  
\]

\[
|k_{2}\rangle = |Fe; d_{xz}\uparrow \downarrow , d_{yz}\uparrow\rangle |As;  p\uparrow \rangle 
\]
Because of the onsite Coulomb repulsion on the d orbital, the state $|k_{2}\rangle $ has s much higher energy than $|k_{1}\rangle $, and it will be ignored. 
We can then write 
\[
\langle 2|H_{eff}^{'}|1\rangle =4\times \frac{1}{2} [ \frac{1}{\epsilon_{2} - \epsilon_{p^{'}} - J}  + \frac{1}{\epsilon_{1}-\epsilon_{p}}] \alpha\beta \equiv \gamma  
\]
 where $J$ is the Hund's exchange term in the d orbitals, $\epsilon_{p}$($\epsilon_{p^{'}}$) is the energy of an 
electron in a p($p^{'}$) orbital on the As atom, and Hund's exchange term in the As 4p orbitals is ignored.
 The factor of 4 results from the fact that an electron on the Fe atom can hop to any of 
the four surrounding As atoms. In a similar way, we can construct the 
remaining matrix elements and diagonalize the Hamiltonian in the 5-dimensional subspace spanned by the states $|1\rangle , ..., |5\rangle$. 

Needless to say, the expressions for the eigenvalues and eigenvectors of the 5x5 matrix are unwieldy. Instead, 
we diagonalize the matrix in the 2-dimensional subspace spanned by $|1\rangle$ and $|2\rangle$, draw the necessary conclusions,
 and then extrapolate to the more general case. 

Thus, assuming for the time being that only states $|1\rangle$ and $|2\rangle$ are mixed by the Hamiltonian, we can express $\mathcal{H_{T}}$ by a  
$2 \times 2$~matrix
\[ \left( \begin{array}{cc}
\epsilon_{1} + \epsilon_{2} - J + 2\epsilon_{p}  & \gamma \\
\gamma  & 2\epsilon_{1} + \epsilon_{p} + \epsilon_{p^{'}} \end{array} \right)\] 
The ground state energy is now given by
\begin{eqnarray*}
E_{gs} &=& \epsilon_{1} + \epsilon_{p} + \frac{1}{2} (\epsilon_{1} + \epsilon_{2} + \epsilon_{p^{'}} + \epsilon_{p} - J )\\ 
       &-& \frac{1}{2} \sqrt{(\epsilon_{1} - \epsilon_{2} + \epsilon_{p^{'}} - \epsilon_{p} + J )^{2} + 4|\gamma|^{2} }
\end{eqnarray*}
and $E_{gs}$ is lower than the energy $E_{1}$ of state $|1\rangle $ by $\delta{E}$ given by
\begin{eqnarray*}
\delta E = E_{gs} - E_{1} &=& \frac{1}{2}[(J-2\Delta + 2\Delta_{p})\\ 
                          &-& \sqrt{ (J-2\Delta + 2\Delta_{p})^{2} + 4|\gamma|^{2} }]  
\end{eqnarray*}
where $2\Delta_{p} = \epsilon_{p^{'}} - \epsilon_{p}$. The unperturbed 
energy $E_{1} = \epsilon_{1} + \epsilon_{2} - J + 2\epsilon_{p} = 2\epsilon + 2\epsilon_{p} - J$ is the same in both the
 tetragonal and orthorhombic phases,
but $\delta{E}$ does depend on the crystalline phase. Since $\gamma$ depends only weakly on $\Delta$, we have 
\begin{eqnarray*}
\delta E_{ortho}^{(e)} - \delta E_{tetra}^{(e)} &=& \delta E(\Delta) - \delta E(\Delta = 0) \\
                                                &=& -\Delta - \frac{1}{2}\sqrt{(J-2\Delta+2\Delta_{p})^{2} + 4|\gamma|^{2}} \\
                                                &+& \frac{1}{2} \sqrt{(J+2\Delta_{p})^{2} + 4|\gamma|^{2}} < 0 \end{eqnarray*}
where the superscript $e$ stands for 'electronic'. The above expression is negative for all values of $J$, $\Delta$ and $\gamma$.
From our earlier discussion, we may take $\epsilon_{d} - \epsilon_{p} \sim0.5$ eV, $\epsilon_{d} - \epsilon_{p^{'}} \sim-1$ eV, 2$\Delta_{p} \sim 1.5$ eV.
 The parameters $J$, $\alpha$, and $\beta$ are all
estimated to be in the range $\sim$0.7-1 eV~\cite{Tao:2008, Cao:2008}; this gives $\gamma \sim1.5-3$ eV.
 The above expression then reduces to 
\[
\delta E_{ortho}^{(e)} - \delta E_{tetra}^{(e)} = -\Delta (1-\frac{J+2\Delta_{p}}{\sqrt{(J + 2\Delta_{p})^{2}+ 4\gamma^{2}}}) + \mathcal{O}(\Delta^{2}) 
\]  
which is negative. We thus conclude that a tetragonal-to-orthorhombic distortion lowers the electronic energy of the d state
 by a term proportional to $\Delta$ 
(ignoring the ${\Delta}^{2}$ term, being smaller than the linear term). The distortion takes place by a small increase in 
 the lattice constant $a$ and a corresponding decrease in lattice constant $b$. Assuming this small distortion is $\delta x $, then $\Delta$
is proportional to $\delta x $, and we can write 
$\delta E ^{(e)} = \xi \delta x $, where $\xi$ is some constant. On the other hand, this distortion raises the elastic 
energy by an amount $\delta E^{L} = \eta (\delta x)^{2}$ where $\eta$ is some constant. The value of the distortion $\delta x$ is found by minimizing 
\[
\delta E^{total} = \delta E ^{(e)} +\delta E^{L} = - \xi \delta x + \eta (\delta x)^{2} 
\] 
giving $\delta x = {\xi}/{2\eta} $, and $ \delta E^{total} = -{\xi^{2}}/{4\eta}$. What we have just shown is that  
 the hybridization between the Fe 3d states and As 4p states 
 causes a greater downward shift in energy if the crystal distorts from tetragonal to 
orthorhombic, and seeking 
the lowest energy available, the crystal will indeed undergo such a distortion. 

The ground state eigenket is obtained from the eigenvector corresponding to the ground state energy, 
\[
|\psi_{gs}\rangle = c_{1}|1\rangle + c_{2} |2\rangle, 
\] 
where
\[
|\frac{c_{2}}{c_{1}}| = \frac{1}{2|\gamma|} [\sqrt{(J + 2\Delta_{p} - 2\Delta )^{2} + 4\gamma^{2}} - (J + 2\Delta_{p}-2\Delta)].
\]
Using the values of the parameters given earlier, we find $|c_{2}/c_{1}|$ in the range $\sim$0.5-0.7.
In the ground state the expectation value of the magnetic moment of the Fe atom is 
\[
\langle \mu \rangle = |c_{1}|^{2} \langle \mu \rangle_{1} + |c_{2}|^{2} \langle \mu \rangle_{2}
\]
Since $\langle \mu \rangle_{2}$, the magnetic moment in state $|2\rangle$, is zero, it follows that $\langle \mu \rangle \sim 0.7\langle \mu \rangle_{1}$. 
In other words, the hybridization causes a large reduction in the value of the would-be magnetic moment. We may now consider the 
mixing of all the states shown in Fig.~\ref{fig:figure2}. The qualitative results obtained so far remain unchanged. Qualitatively, the ground state 
will be a mixture of five states, in four of which the magnetic moment is zero. Therefore, we expect a further reduction in the expectation value of the Fe 
magnetic moment. In fact we do not need to restrict ourselves to these states; we may also consider mixing with states where one electron resides on the 
d$_{x^{2}-y^{2}}$ orbital. All this leads to the conclusion that strong hybridization between the Fe 3d and As 4p orbitals causes 
spin fluctuations on the Fe sites and a subsequent reduction in the measured value 
of the Fe magnetic moment, in agreement with neutron scattering measurements.~\cite{Zocco:2008, Ren:2008, Chen:2008, x_Chen:2008}

We now consider what happens as some oxygen atoms are replaced by fluorine atoms. Fluorine doping causes further 
negative charge transfer from the ReO/F layer to the As atoms, since it involves replacing O$^{2-}$ with F$^{-}$. This strengthens the ionic bond and 
shortens the distance between the Re and As ions. Indeed, neutron scattering measurements ~\cite{Cruz:2008} indicate that 
the La-As separation decreases by 0.03 $\AA$ upon F-doping, with a concomitant reduction in the c-lattice constant. But the extra charge,
donated by the Re ions, is not uniformly
spread among the As ions; more of this charge resides on As ions close to the F atoms, rather than
on those further away, and some local distortion of the As ion positions is expected.
With such distortion, and with an uneven charge distribution on the As ions, the crystal field at a given Fe site 
now has a symmetry lower than $S_{4}$. Considering a single Fe atom, the Hamiltonian is written as 
\[
\mathcal{H} = \mathcal{H}_{0} + V 
\]
where $\mathcal{H}_{0}$ contains the part of the crystal field with S$_{4}$ symmetry, while V is the additional part which lowers that symmetry. 
The orbitals Fe 3d$_{xz}$ and 3d$_{yz}$ are eigenstates of $\mathcal{H}_{0}$ but not $\mathcal{H}$; the perturbation V mixes these two states.
Let $\langle 3d_{xz}|V|3d_{xz}\rangle = V_{11} $, $\langle 3d_{yz}|V|3d_{yz}\rangle = V_{22} $, and $\langle 3d_{xz}|V|3d_{yz}\rangle = V_{12} $;
 these matrix elements are proportional to the F-dopant concentration. 
Then the energy splitting 2$\Delta$ of the two states is given by
\[
2\Delta = \sqrt{(V_{11} -V_{22})^{2} + 4V_{12}^{2} } ,
\]
and the two new split states $|a\rangle$ and $|b\rangle$ are linear combination of $|3d_{xz}\rangle$ and $|3d_{yz}\rangle$.
As the F dopant concentration increases, the energy splitting increases, and if 2$\Delta$ becomes larger than $J$, the contribution
of state $|2\rangle$, shown in Fig~\ref{fig:figure2}, to the ground state wave function becomes dominant,
 and the Fe magnetic moment becomes very small, causing the magnetic order to be suppressed.
If the fluorine concentration is $x$, then in the ReO/F layer, every position in the O/F plane is occupied by a fluorine atom
with probability $x$ and by an oxygen atom with probability $1-x$. The overall symmetry of the crystal structure, as seen in 
X-ray measurements, is still tetragonal with symmetry group P4/nmm. The instantaneous potential seen by an electron depends on the 
ionic positions and is given by
\[
V= V_{s} + \sum_{k} ({\partial{V}}/{\partial{Q_{k}}}) Q_{k} + ... 
\]
where V$_{s}$ is the static part of the potential, Q$_{k}$ is a lattice normal coordinate, and because of the invariance of V 
under the crystal symmetry group,  ${\partial{V}}/{\partial{Q_{k}}}$ transforms according to the same irreducible representation as Q$_{k}$. Since in
the presence of F doping, the Fe electronic orbitals $|i\rangle$ are nondegenerate, then $\langle i| {\partial{V}}/{\partial{Q_{k}}}|i\rangle$ 
vanishes unless ${\partial{V}}/{\partial{Q_{k}}}$ is totally symmetric under the crystal symmetry group operations. It follows that any 
spontaneous distortion that occurs must be totally symmetric and will not lower the crystal symmetry. Hence, F doping suppresses the 
tetragonal-to-orthorhombic distortion. 

In conclusion, we have shown that strong hybridization between the Fe 3d orbitals and the As 4p orbitals
 introduces some interesting effects. In the undoped systems, 
it causes a larger downward shift in the electronic energy if the crystal undergoes a tetragonal-to-orthorhombic distortion which lifts the degeneracy 
of the Fe 3d$_{xz}$ and 3d$_{yz}$ levels. The ground state, in the presence of hybridization, is found to be a mixture of states, most of them with 
zero Fe magnetic moment, so that in this ground state the expectation value of the Fe magnetic moment is small.
 We have also argued that the symmetry lowering caused by 
fluorine doping suppresses both the lattice distortion and the magnetic order of the Fe moments.
 Finally, we note that although the mechanism for superconductivity
in the iron-based superconductors is not understood yet, we believe that the spin fluctuations on the Fe sites, caused by hybridization
between the Fe 3d and the As 4p states, may play a role, not only in suppressing the magnetic order, but also in the emergence of
superconductivity in these compounds.

We gratefully acknowledge useful discussions with Professor M. S. Dresselhaus.


\end{document}